# Oxide layer dependent orbital torque efficiency in ferromagnet/Cu/Oxide heterostructures

Junyeon Kim[1], Jun Uzuhashi[2], Masafumi Horio[3], Tomoaki Senoo[3], Dongwook Go[4,5], Daegeun Jo[6], Toshihide Sumi[3], Tetsuya Wada[3], Iwao Matsuda[3,7], Tadakatsu Ohkubo[2], Seiji Mitani[2], Hyun-Woo Lee[6], and YoshiChika Otani[1,3,7]

[1]*Center for Emergent Matter Science, RIKEN, Wako, Saitama, 351-0198, Japan*

[2]*National Institute for Materials Science, Tsukuba, Ibaraki, 305-0047, Japan*

[3]*Institute for Solid State Physics, The University of Tokyo, Kashiwa, Chiba, 277-8581, Japan*

[4]*Peter Grünberg Institut and Institute for Advanced Simulation, Forschungzentrum Jülich and JARA, 52428 Jüllich, Germany*

[5]*Institute for Physics, Johannes Gutenberg University, Mainz, 55099 Mainz, Germany*

[6]*Department of Physics, Pohang University of Science and Technology, Pohang 37673, South Korea*

[7]*Trans-scale Quantum Science Institute, The University of Tokyo, Tokyo, Japan*

**The utilization of orbital transport provides a versatile and efficient spin manipulation mechanism. As interest in orbital-mediated spin manipulation grows, we face a new issue to identify the underlying physics that determines the efficiency of orbital torque (OT). In this study, we systematically investigate the variation of OT governed by orbital Rashba-Edelstein effect at the Cu/Oxide interface, as we change the Oxide material. We find that OT varies by a factor of ~2, depending on the Oxide. Our results suggest that the active electronic interatomic interaction (hopping) between Cu and oxygen atom is critical in determining OT. This also gives us an idea of what type of material factors is critical in forming a chiral orbital Rashba texture at the Cu/Oxide interface.**

Since the early reports on the successful magnetization switching induced by charge-to-spin conversion [1,2], intensive studies to find new physics and materials for an efficient switching have been carried out for a decade [3]. Although large charge-to-spin conversion efficiency has been achieved, the rather narrow range of material selection restricted to heavy elements (e.g. Pt, Bi) is not favorable for versatile and functional applications. Recently, an alternative approach of spin manipulation by the orbital angular momentum (OAM) has opened novel opportunities [4,5]. Similar to the charge-to-spin conversion, the non-equilibrium OAMs (charge-to-OAM conversion) can be generated at the interface/surface by orbital Rashba effect (ORE)/orbital Edelstein effect (OEE) or in bulk systems by orbital Hall effect (OHE) [6-8]. Remarkably, both ORE/OEE and OHE do not resort to spin-orbit coupling, thus they could be significant in light element material systems. If the generated non-equilibrium OAMs are transferred into a ferromagnet (FM), they exert a torque by the interaction with spins in the FM (orbital torque (OT)), and subsequently alter the magnetization orientation [8]. The orbital torque is predicted to be considerable even though the spin-orbit coupling of an FM is restricted since the charge-to-OAM conversion is immensely efficient in diverse material systems as theoretically predicted to be 10-100 times larger than the conventional charge-to-spin conversion [9,10]. Considerable OT was experimentally verified in various light material systems such as Cu/$Al_2O_3$ interface, naturally oxidized Cu surface, and bulk light metals (e.g. Cr, and Ti) [11-19]. Notably, the efficiency of the OT is energetically 10-100 times higher than the spin torque derived by the conventional charge-to-spin conversion [11]. Recent reports reveal that injection of non-equilibrium OAMs together with spins also provides a reliable mechanism for the spin manipulation [14,20-22]. Successful experimental verifications of magnetization switching and magnetoresistance due to OAM injection imply that the charge-to-OAM conversion could provide an analogous impact to the charge-to-spin conversion [11,15,23].

The emergence of the OAM utilization naturally gives rise to a question: what kind of material factors determine the efficiency of spin manipulation. Since theoretical predictions indicated an extremely large efficiency of the OHE [7,10], experimental reports have verified the prediction in various light bulk metals. From these studies, another interesting issue on long-range orbital transport is raised [15-17]. On the other hand, there are few report discussing the key factors for ORE- and OEE- induced non-equilibrium OAMs generation such as theoretical study on the oxygen-incorporation role at Cu surface for ORE [9]. In this letter, we report a systematic experimental study on the material dependence of OT in $Co_{25}Fe_{75}$ (CoFe)/Cu/Oxide structures. Observed OT efficiencies in various CoFe/Cu/Oxide structures are generally considerable regardless of the Oxide, revealing the material generality of ORE/OEE. Despite the generality, we also observe a large Oxide-dependent variation of OT with a factor of ~2. Our analysis based on the Oxide dissociation indicates that the oxygen atom release from the Oxides affects the OT most crucially than other physical properties of the oxide do.

We prepare CoFe (5)/Cu (0-30)/Oxide (=MgO, $SiO_2$, $TiO_2$, 20) (nm) heterostructures using the electron-beam evaporation technique (Fig. 1(a)). For comparison, we also included OT results from CoFe (5)/Cu (0-30)/$Al_2O_3$ (20) (nm) from Ref. [11] for analysis. In this letter, we call them MgO, $SiO_2$,

TiO$_2$, and Al$_2$O$_3$ samples, respectively, depending on the Oxide in the CoFe/Cu/Oxide structures. The Cu layer is prepared in a wedged layer by utilizing the linear shutter during the deposition. We patterned devices for spin torque ferromagnetic resonance (ST-FMR, Fig. 1(b)) and spin/OAM pumping measurement using photolithography and lift-off process. The dimensions of the devices for ST-FMR and spin/OAM pumping measurement are 20 μm×60 μm, and 8 μm×200 μm, respectively.

We evaluate the efficiency of the OT using the ST-FMR technique. The amplitude of the rectified resonance signal under the application of the external field and injection of the GHz range OAM (spin) current is a measure of the orbital (spin) torque [24]. The resonance signal is decomposed to symmetric $S$ and anti-symmetric $A$ component Lorentzian functions, amplitude of which are proportional to antidamping torque and addition of fieldlike torque and Oersted field torque, respectively. The ST-FMR spectrum for the MgO, SiO$_2$, and TiO$_2$ samples is shown in Fig. 1 (c)-(e), and a clear peak of the symmetric component of Lorentzian is observed, indicating considerable exertion of antidamping torque. The height of the peak is found to differ depending on the Oxide, suggesting that the Oxide plays an important role in determining the ORE. We note that the large symmetric Lorentzian peak disappears when we replace CoFe to Ni$_{80}$Fe$_{20}$ (Py) (Supplemental information Fig. S1). It is consistent with the previous reports that the OT disappears when Py is adopted as a FM [11,12,16]. We suspect that unintended scattering due to the intermixing between Ni and Cu may critically reduce the orbital transparency at the Py/Cu interface [8,11]. The noticeable FM dependence supports that the torque in the CoFe/Cu/Oxide structures would come from the orbital transport. We also note that spin/OAM-to-charge conversion examined by the spin/OAM pumping has extremely low efficiency (Supplemental information Fig. S2), contrasting with the considerable torque exertion. Such disaccordance of the efficiencies was reported in two-dimensional systems [11,12,25,26]. As a result, we thus do not need to consider a problematic charge current generation by the spin/OAM pumping during the ST-FMR measurement [27].

The Cu thickness dependence of OT is studied. The efficiency of OT ($\theta$) is evaluated by the following equation,

$$\theta = \frac{S}{A} \frac{4\pi M_S e t_{CF} d_{Cu}}{\hbar} \left[1 + \frac{4\pi M_{eff}}{H_{ext}}\right]^{1/2}, \qquad (1)$$

$4\pi M_S$ and $4\pi M_{eff}$ the saturation magnetization and the effective saturation magnetization of the CoFe layer. $e$ the elementary charge, $t_{CF}$ the CoFe layer thickness, $d_{Cu}$ the Cu layer thickness, $\hbar$ the reduced Planck constant, and $H_{ext}$ the external magnetic field.

The Cu thickness dependence of the OT for all the structures is presented in Fig. 2(a). Note that the OT results for the Al$_2$O$_3$ sample comes from our previous study in Ref. [11]. We find that the variation of $\theta$ by the Cu thickness is similar for all structures. Specifically, $\theta$ increases with increasing Cu thickness in the thin Cu thickness regime, and reaches maximum at a certain Cu thickness. After that, $\theta$ decreases with an increase in the Cu thickness in thick Cu thickness regime. This Cu thickness behavior is widely observed in interface/surface-originated orbital (spin) torque and OAM (spin)-to-spin conversion in various material structures [11,12,26,28,29]. In these samples, several factors compete to determine $\theta$. One factor is the crystallinity of the Cu layer and the Cu/Oxide interface, which would improve with an increase in Cu thickness. Better crystal-ordered Cu/Oxide interface would strengthen the ORE, resulting in increasing $\theta$ with the Cu thickness. On the other hand, the current shunting to the Cu layer and decay of OAM during transport bring about a decreasing $\theta$ with the Cu thickness. The competition among these factors can generate the non-monotonic Cu thickness dependence of $\theta$, as depicted in Fig. 2(a). Together with the non-reciprocity of the efficiencies mentioned above, the Cu thickness dependence of $\theta$ shown in Fig. 2(a) indicates that the interfacial ORE/OEE is the dominant mechanism to produce OT for these structures. In addition to the qualitative similarity among various Oxide samples, the maximum value of $\theta$ for each sample is sizable regardless of the Oxide choice although there is two-fold difference: $\theta_{max}$~0.26, 0.24, 0.17 and 0.12 for the $SiO_2$, $TiO_2$, MgO and $Al_2O_3$ samples, respectively. Moreover the Cu thickness for $\theta_{max}$ is also similar ($d_{Cu}$~15 nm) for the $SiO_2$, $TiO_2$, and MgO samples, though the Cu thickness for $\theta_{max}$ is smaller ($d_{Cu}$~10 nm) for the $Al_2O_3$ sample.

To understand the origin of the OT better, we examine physical factors that may be responsible for the Oxide dependence of $\theta_{max}$. We first consider the dielectric constant of the Oxide as a possible candidate, since the ORE relies on the structural inversion asymmetry (SIA) and non-uniform electron distribution at the surface/interface [6,30,31]. We hypothesized that a higher dielectric constant could enhance the non-uniformity if the electric field by the SIA is considerable. However, our experimental results does not show a clear correlation between $\theta_{max}$ and the dielectric constant of the Oxide (Supplemental information Fig. S3).

Next, we consider the interaction between the Cu and oxygen atoms. As briefly mentioned earlier, the recent theoretical study reported that active electronic hopping between a *p*-orbital of an oxygen atom and *d*-orbitals of a Cu atom (orbital hybridization) critically promotes the SIA at an oxygen-incorporated Cu surface [9]. The promotion of the SIA eventually strengthens the ORE and assists efficient OAM polarization. Additionally, the orbital hybridization and the ORE are strongly enhanced as the distances between constituent atoms shorten [32]. In this line, we propose that a large number of oxygen atoms released from an Oxide might increase the number of Cu atoms that interact with oxygen atoms and/or shorten the distance between Cu and oxygen atoms. These factors should promote the ORE. We thus attempt to estimate the number of released oxygen atoms ($N_R$) by considering the oxygen binding energy ($E_b$) and the possible oxygen supply number ($N_0$) of an Oxide. We adopt Gibbs free energy as a measure of $E_b$ since this parameter considers all possible physical issues (e.g. change of the entropy and the volume) during a reaction. Here, small (large) $|E_b|$ means an unstable (stable) state (Supplemental information Table S1). Among the Oxides, $Al_2O_3$ (MgO) is the most stable (unstable). And we define $N_0$ as the number of oxygen atoms per each cation (i.e. $A_XO_Y$ with a cation A, then $N_0$=Y/X). In order to obtain large $N_R$, smaller $|E_b|$ and larger $N_0$ are preferred, indicating more oxygen atoms contained in an unstable Oxide. We thus estimate $N_R$ by the relation $N_R = N_0/|E_b|$. The correlation between $\theta$ and $N_R$ is evident as shown in Fig. 2(b), where small (large) $\theta$ is obtained when $N_R$ is small (large), as demonstrated in the case of the $Al_2O_3$ ($SiO_2$ or $TiO_2$) sample. This correlation stresses that the interatomic interaction between Cu and the oxygen atoms is a critical factor in determining the OT. Furthermore, it suggests the main mechanism for the formation of orbital Rashba texture at the Cu/Oxide interface is the orbital hybridization between Cu and oxygen atoms. It also provides a possible explanation as to why $\theta_{max}$ is achieved at a larger Cu thickness for the $SiO_2$, $TiO_2$ and MgO samples. With more release, oxygen atoms could permeate the Cu layer deeper, resulting in a shorter pristine Cu region and weakening the decay of the OAM in a thick Cu layer. To some extent, this analysis supports the idea that the SIA by the electronic hopping plays a dominant role in the formation of the orbital Rashba texture at the Cu/Oxide interface, rather than the generation of a uniform surface electric field at there [32,33].

In order to verify the analysis above, we carried out the cross-sectional high-angle annular dark-field scanning transmission electron microscope (HAADF-STEM) observation with the energy dispersive X-ray spectroscopy (EDS) analysis for the $Al_2O_3$ and $SiO_2$ sample (Fig. 3). All the EDS line concentration profiles are averaged value for 35 nm-width regions (Fig. 3(a)). The observed EDS line concentration profiles near the Cu/Oxide interface for the $Al_2O_3$ and $SiO_2$ samples are displayed in Fig. 3(c) and (d), respectively. We select the oxygen line profile for these samples for direct comparison (Fig. 3 (b)). We find that the oxygen line profile for the $Al_2O_3$ sample has a relatively steep slope, while the $SiO_2$ sample has a relatively gentle slope. This suggests that more oxygen atoms are released and could permeate the Cu layer deeper in the $SiO_2$ sample, which is consistent with our analysis in Fig. 2(c). Hence, we presume that more Cu atoms could interact with oxygen atoms in the $SiO_2$ sample. By the way, we do not find a noticeable difference in the slope of the oxygen accumulation at the CoFe/Cu interface between the $Al_2O_3$ and $SiO_2$ samples (Supplemental information Fig. S4). Thus the difference in OT between the $Al_2O_3$ and $SiO_2$ samples could not be explained by an additional influence from the modulation of the orbital Rashba texture at the CoFe/Cu due to the oxygen accumulation [34].

Additionally, we carried out the X-ray absorption spectroscopy (XAS) observation for the Cu/Oxide interface [35,36]. Since the X-ray can penetrates the transparent oxide layer but not the opaque Cu layer, hence the XAS results presented in this letter mainly reflect material information near the Cu/Oxide interface. Figure 4 displays the XAS spectra for the $Al_2O_3$, $SiO_2$, and $TiO_2$ samples. The XAS spectrum for the $Al_2O_3$ sample confirms the presence of only metallic Cu at the $Cu/Al_2O_3$ interface (Fig. 4(a)). On the other hand, we observe the formation of Cu oxide (CuO or $Cu_2O$) at the Cu/Oxide interface in the XAS spectrum for the $SiO_2$ (Fig. 4(b)), and $TiO_2$ sample (Fig. 4(c)). We speculate that the released oxygen atoms could form Cu oxides among the larger pool of released oxygen atoms. Thus, the XAS spectra also support the above mentioned argument for the interatomic interaction between Cu and oxygen atoms.

In summary, we carried out the OT measurement for several CoFe/Cu/Oxide samples, and we found that a considerably large OT is observed for all the material systems, which verifies the material generality of ORE and OT. The magnitude of OT has a clear correlation with the oxygen release of the Oxide, supporting the argument that the interatomic interaction between Cu and the oxygen atom is a critical factor for the generation and magnitude determination of OT. This argument is also supported by material characterization observation studies.

We are grateful to K. Hono and L. Liao for the fruitful discussion. J.K. and Y.O. appreciate the financial support from JSPS KAKENHI (Grant No. 19K05258 and 19H05629). D.J. and H.-W.L. were supported by the SSTF Foundation (BA-1501-51). The synchrotron radiation experiments were performed at the BL07LSU of SPring-8 with the approval of the Japan Synchrotron Radiation Research Institute (JASRI) (Proposal No. 2021B7410, 2022A7411, and 2022A7412).

**Figure captions**

**Fig. 1.** (a) Scheme of the CoFe/Cu/Oxide structures. (b) Scheme of the ST-FMR device. (c), (d), (e) ST-FMR spectrum for the MgO (c), SiO$_2$ (d), and TiO$_2$ (e) sample. All the samples have 15 nm of Cu thickness. 14 dBm power and 11 GHz frequency of input was injected. Red dots are the raw data. Blue curves, green curves, and red curves represent the symmetric Lorentzian, asymmetric Lorentzian, and addition of both, respectively.

**Fig. 2.** (a) OT as a function of Cu thickness for all the samples. Yellow up-pointing triangles, blue circles, red squares, and green down-pointing triangles represent OT from the SiO$_2$, TiO$_2$, MgO and Al$_2$O$_3$ samples, respectively. Note that OT results for the Al$_2$O$_3$ sample come from Ref. [11]. (b) Maximum OT ($\theta_{max}$) for all the structures as a function of the number of released oxygen atoms $N_R$.

**Fig. 3**. (a) STEM images for the Al$_2$O$_3$ samples, representing the observed area for the EDS analysis. (b) Oxygen atom EDS line profile near the Cu/Oxide interface for the Al$_2$O$_3$ (red line) and SiO$_2$ (blue line) samples. (c), (d) EDS line concentration profiles near the Cu/Oxide interface for the Al$_2$O$_3$ (c), and SiO$_2$ (d) samples. Blue, green, brown, and yellow lines represent the composition ratio of oxygen, Al, Cu, and Si atoms, respectively. For all the graphs for the EDS analysis, Oxide (Cu) layer is located when value of the position closes to 0 (8) nm.

**Fig. 4.** (a), (b), (c) XAS spectrum for the Al$_2$O$_3$ sample (a), SiO$_2$ (b), and TiO$_2$ (c) samples. Guide lines for (b) and (c) represents the position of peak for CuO (left) and Cu$_2$O (right).

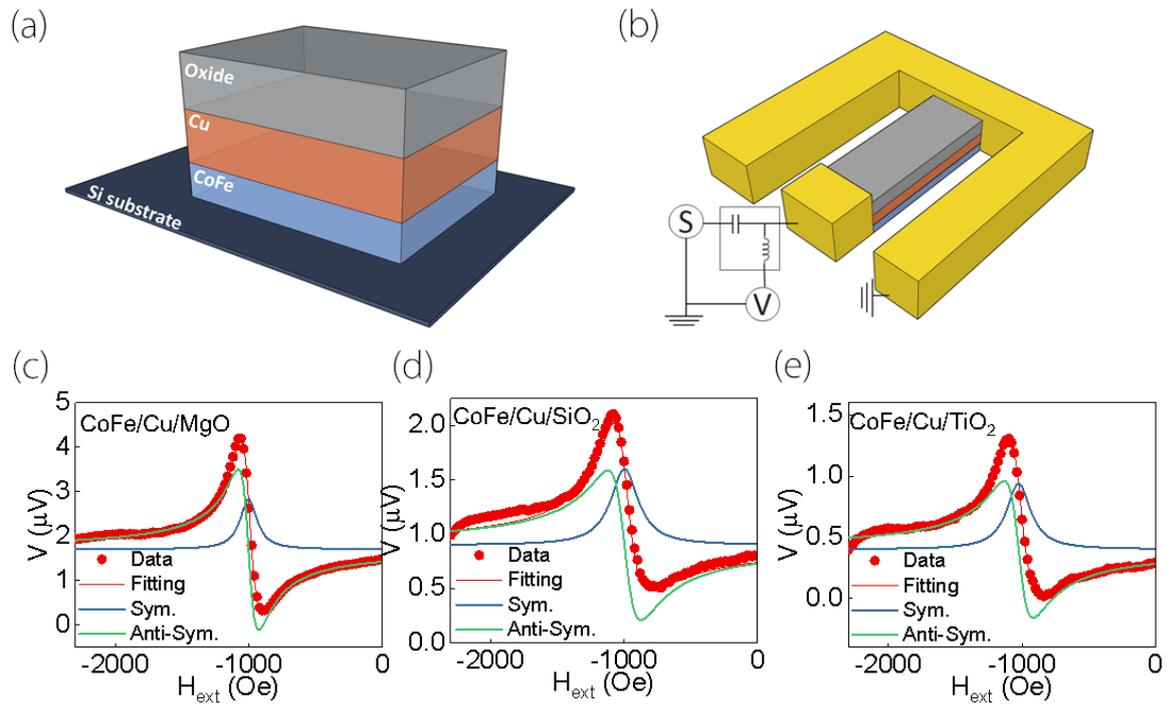

**Fig. 1.** (a) Scheme of the CoFe/Cu/Oxide structures. (b) Scheme of the ST-FMR device. (c), (d), (e) ST-FMR spectrum for the MgO (c), SiO$_2$ (d), and TiO$_2$ (e) sample. All the samples have 15 nm of Cu thickness. 14 dBm power and 11 GHz frequency of input was injected. Red dots are the raw data. Blue curves, green curves, and red curves represent the symmetric Lorentzian, asymmetric Lorentzian, and addition of both, respectively.

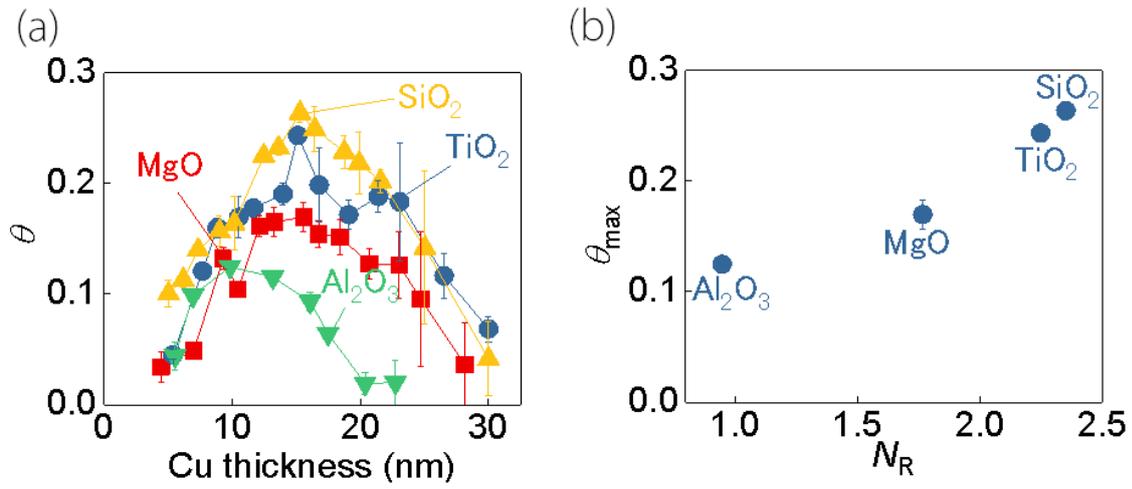

**Fig. 2.** (a) OT as a function of Cu thickness for all the samples. Yellow up-pointing triangles, blue circles, red squares, and green down-pointing triangles represent OT from the $SiO_2$, $TiO_2$, MgO and $Al_2O_3$ samples, respectively. Note that OT results for the $Al_2O_3$ sample come from Ref. [11]. (b) Maximum OT ($\theta_{max}$) for all the structures as a function of the number of released oxygen atoms $N_R$.

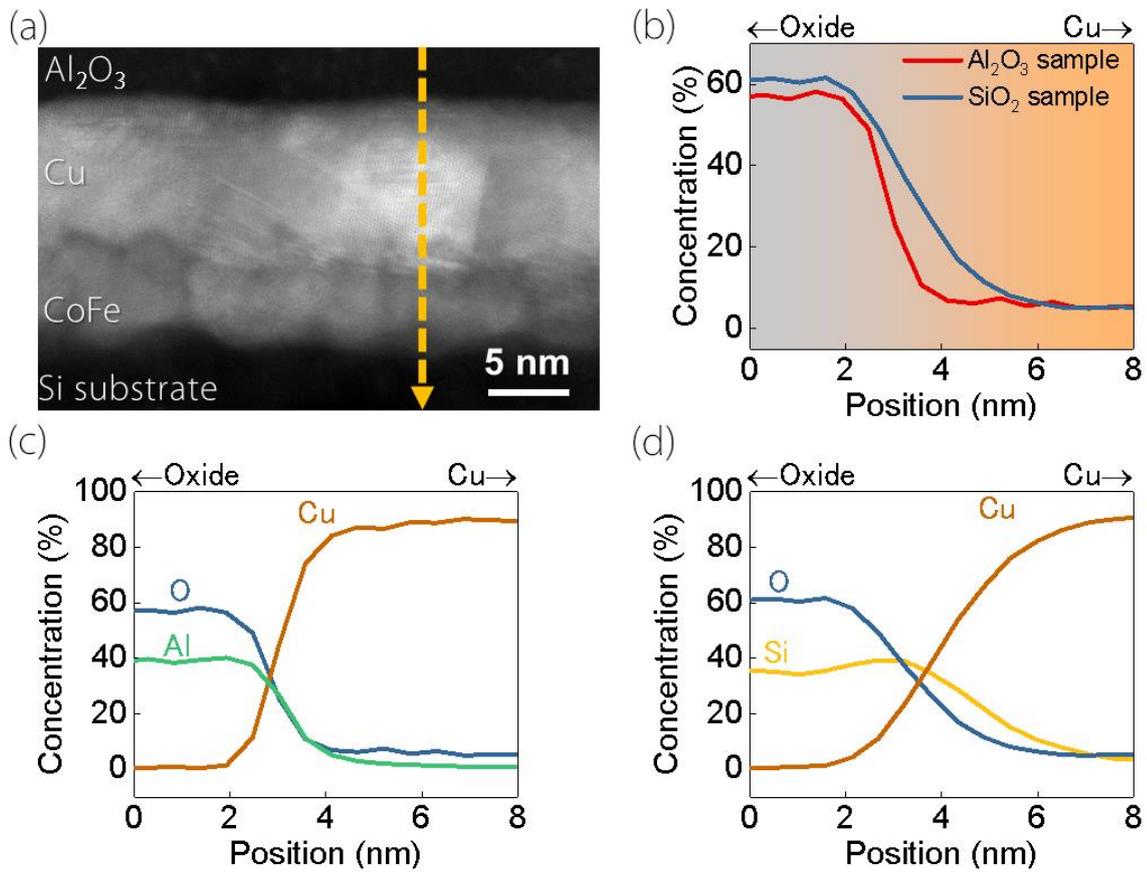

**Fig. 3**. (a) STEM images for the $Al_2O_3$ samples, representing the observed area for the EDS analysis. (b) Oxygen atom EDS line profile near the Cu/Oxide interface for the $Al_2O_3$ (red line) and $SiO_2$ (blue line) samples. (c), (d) EDS line concentration profiles near the Cu/Oxide interface for the $Al_2O_3$ (c), and $SiO_2$ (d) samples. Blue, green, brown, and yellow lines represent the composition ratio of oxygen, Al, Cu, and Si atoms, respectively. For all the graphs for the EDS analysis, Oxide (Cu) layer is located when value of the position closes to 0 (8) nm.

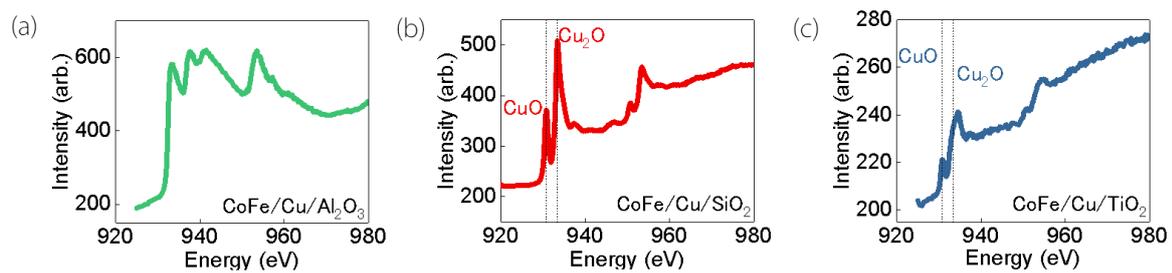

**Fig. 4.** (a), (b), (c) XAS spectrum for the $Al_2O_3$ sample (a), $SiO_2$ (b), and $TiO_2$ (c) samples. Guide lines for (b) and (c) represents the position of peak for CuO (left) and $Cu_2O$ (right).